# A SEARCH FOR DOUBLE-LOBED RADIO EMISSION FROM GALACTIC STARS AND SPIRAL GALAXIES

Abiel Felipe Ortiz Martínez[1] and Heinz Andernach[2]

## RESUMEN

Se presenta una búsqueda sistemática de dos tipos de objetos astronómicos con inusuales características: estrellas y galaxias espirales, con dobles lóbulos en radio, lo cual sugiere la eyección de chorros desde el objeto óptico. Diseñamos un algoritmo para buscar pares de radiofuentes en lados opuestos de dos muestras inéditas de 878,031 estrellas del *Sloan Digital Sky Survey* y 675,874 galaxias candidatas a espirales extraídos de la literatura. Se encontraron tres nuevos ejemplos de radio estrellas con fuentes dobles, mientras que para galaxias espirales sólo se redescubrió una fuente doble conocida, confirmando que estos últimos son extremadamente raros.

## ABSTRACT

We present a systematic search for two types of very unusual astronomical objects: Galactic stars and spiral galaxies with double radio lobes, i.e. radio emission on opposite sides of the optical object, suggesting the ejection of jets from them. We designed an algorithm to search for pairs of radio sources straddling objects from two unprecedented samples of 878,031 Galactic stars from the *Sloan Digital Sky Survey* and 675,874 spiral galaxy candidates drawn from the recent literature. We found three new examples of double-lobed radio stars, while for the spiral galaxies we only rediscovered one known such double source, confirming that the latter objects are extremely rare.

**Palabras Clave**: astrofísica extragaláctica, galaxias espirales, radioestrellas, radiogalaxias

## INTRODUCTION

Double-lobed radio sources that originate either from stars or from spiral galaxies are extremely rare. We thus decided to perform a new search for more examples of these types of objects.

Within the citizen science project "Radio Galaxy Zoo" (RGZ, radio.galaxyzoo.org), launched in 2013, and of which one of us (HA) is a founding member, several spectroscopically confirmed Galactic stars were found to lie on the axis connecting two radio sources. These suspected "radio lobes" are often extended along the axis between the lobes, very similar to the morphology of the radio emission of classical double radio galaxies. However, different from the latter, the stars usually do not present a radio nucleus at the position of the star. The discovery by a RGZ volunteer of a V~9 mag star (BD+40 2030) motivated one of us (HA) to search the best available radio source catalogue (FIRST; Helfand, White & Becker 2015) performed at 1.4 GHz with an angular resolution of 5.4´´ for double radio sources straddling the position of V<11 mag stars within ~30´´, finding a total of five such cases.

The only known type of double-lobed radio stars are the so-called microquasars (Mirabel & Rodríguez 1999), which are stars that have dark companions (black holes or neutron stars) causing the ejection of radio lobes through mass donation from the visible star onto the dark companion, creating an accretion disk around the latter. All of these are close to the Galactic plane and are copious X-ray emitters, very unlike the double-lobed radio stars found by us to coincide with spectroscopically normal, high Galactic latitude, and X-ray quiet stars. Three of the latter stars have been followed up in a spectroscopic monitoring program at TIGRE observatory (Schmitt et al. 2014) to search for possible dark companions, but so far no radial velocity changes were seen, suggesting that any such companion must be in a much wider orbit with a much longer orbital period. In the present work we extend the search for such stars to fainter magnitudes, by searching for radio emission different from point sources (a type of emission known for several decades to occur in certain types of active stars). Previous searches for radio-emitting stars in the FIRST radio survey were either restricted to brighter stars, and did not reveal any radio double sources (Helfand et al.,1999), or to fainter stars (i-band magnitude from 15 to 19.1 mag), reporting no detections of point sources in excess of chance coincidence, but another 76 stars with complex radio emission (Kimball et al., 2009). For the unresolved sources these authors restricted their radio source search within a separation of at most 1´´ from the star, while for the more complex radio sources they visually inspected an area of 2´ x 2´ around these.

[1] Universidad Autónoma de San Luis Potosí, Facultad de Ciencias, Álvaro Obregón #64, Col. Centro, C.P. 78000, San Luis
[2] Universidad de Guanajuato, Departamento de Astronomía, DCNE, Cjón. de Jalisco s/n, Col. Valenciana, C.P: 36240, Guanajuato, GTO, heinz@astro.ugto.mx

On the other hand, the so-called "classical" radio galaxies are almost exclusively hosted by massive, gas-poor elliptical galaxies (e.g. Best et al. 2005). Only since the work of Ledlow, Owen & Keel (1998) about half a dozen spiral galaxies have been found (see Mao et al. 2015 for a review). Recently Singh et al. (2015) searched for extended radio sources within 3´ on opposite sides of 187,005 spiral galaxies which could indicate double radio lobes ejected from their nuclei. They found only four examples, three of which had been known before. In the present work we apply a similar search algorithm to a 3.6 times larger sample of spirals, and without the restrictions imposed by Singh et al. (2015), corresponding in effect to a more profound search by a few tens of times.

## METHODS AND MATERIALS

A. The Search Algorithm

For both of the searches, around stars and around spiral galaxies, we adapted a FORTRAN code originally written by M. A. Jiménez Valencia (Univ. de Sonora, Mexico) during a summer research project with one of us (HA) in 2015. Our code uses all radio sources in a given field, and, for each pair of radio sources, it calculates (a) the angular distances from the central star or galaxy, (b) the orientation of the radius vector from the central object to the radio source (the position angle), (c) the so-called "armlength ratio" (ALR) of the distance between central object and the radio sources on each side, (d) the difference between the two position angles (the so-called "misalignment" or bending angle, or DPA), and (e) the ratio of their integrated flux densities as listed in the FIRST catalog (the so-called flux ratio, FLR). Only source pairs which lie on opposite sides of the central object, within a given tolerance of armlength ratio, bending angle and flux ratio were kept for further inspection. In our initial inspection we concentrated on the most symmetric and aligned objects, with ALR and FLR closest to unity and DPA closest to 0°.

B. Stars

For our search for double-lobed radio stars we used all 998,703 objects of the *Sloan Digital Sky Survey* Data Release 12 (SDSS DR12, Alam et al. 2015) classified as Galactic stars. Using TOPCAT (Taylor 2013) we kept only one of all duplicate objects within 1´´, leaving us with 878,031 stars. For the current exploratory search, and based on previous experience with bright stars, we restricted our search to radio sources within 1´ (i.e. 60´´) from the stars, and extracted radio sources from the FIRST catalog (Helfand, White, & Becker, 2015) containing 946,432 radio sources.

C. Spiral Galaxies

Samples of spiral galaxies were extracted from four different sources. Firstly, Kuminski & Shamir (2016, KS16) applied an automated algorithm to classify ~3,000,000 galaxies. From their spectral sample we selected those classified as spiral (S), having redshift z > 0.003, a probability of being a star of $p_{star}$<0.01, and a probability of being a spiral of $p_{spi}$>0.80. Inspecting SDSS images (http://skyserver.sdss.org/DR13//en/tools/explore/summary.aspx) led us to select 170,823 further objects with 0.6<$p_{spi}$<0.8, $p_{star}$<0.1, and z>0.002. No upper redshift limit was imposed since we found several supposedly high-z objects that are in fact faint low-surface brightness objects with wrongly determined redshift. From the non-spectral sample we selected all objects with $p_{spi}$ > 0.6. A cleaning of all duplicate objects within 1.5" using TOPCAT resulted in 575,486 spirals. Secondly, Willett et al. (2013) have used the votes of over 87,000 volunteers of the Galaxy Zoo 2 (GZ2) project to classify 304,122 galaxies bright enough to allow morphological classification from visual inspection. We considered all galaxies whose "most common consensus classification", `gz2_class`, started with the letter S, indicating any sub-type of a spiral galaxy. After removal of identical objects contained in different subsamples of GZ2 this resulted in 167,249 spiral galaxy candidates. Thirdly, Huertas-Company et al. (2011, HC11) used support vector machines to classify 698,420 galaxies into 4 morphological types (E, S0, Sab, Scd). They also list the "Automatic Spectroscopic K-means based (ASK) classification" as derived by Sánchez-Almeida et al. (2010) from optical spectra of these galaxies. We selected all galaxies with a probability of at least 0.7 for being Sab or Scd. From the remaining galaxies we added to our sample those with ASK > 14, which according to Sánchez Almeida et al. (2010) is a good indicator of a late-type spiral. We then excluded all objects with probabilities > 0.5 of being either an E/S0 or S0 type, leaving 129,527 galaxies. Finally, to complete our spiral galaxy sample for the brightest objects, we selected from HyperLEDA (http://leda.univ-lyon1.fr) all galaxies with btc < 16 mag, morphological type T>2.0 and ΔT<0.5, resulting in 1150 galaxies in the sky area covered by the FIRST survey. Excluding duplicates within 2.5" of the combined KS16, GZ2,

HC11, and HyperLEDA samples, we ended up with 675,874 spiral candidates. Our final sample is thus ~3.6 times larger than the sample size of Singh et al. (2015). Moreover, we did not limit ourselves to spirals having a detected radio core and searched an area around these objects four times as large, corresponding in effect to a search of a few tens of times larger than the largest previous systematic search.

**RESULTS**

A. Stars

Using TOPCAT we extracted all radio sources from the FIRST catalog within 1´ of 878,031 stars, keeping only those 7192 stars with two or more such radio sources. Applying our algorithm to these 16,693 radio sources we found 524 radio source pairs with a high level of symmetry. Initially we included all objects with ALR >10 to find possibly one-sided radio sources for which one source coincided with the star and the other may be ejected from the star. The few cases we had time to inspect did not show evidence for a one-sided ejection, but we confirmed about a dozen unresolved radio sources coinciding with stars that had already been identified by Kimball et al. (2009).

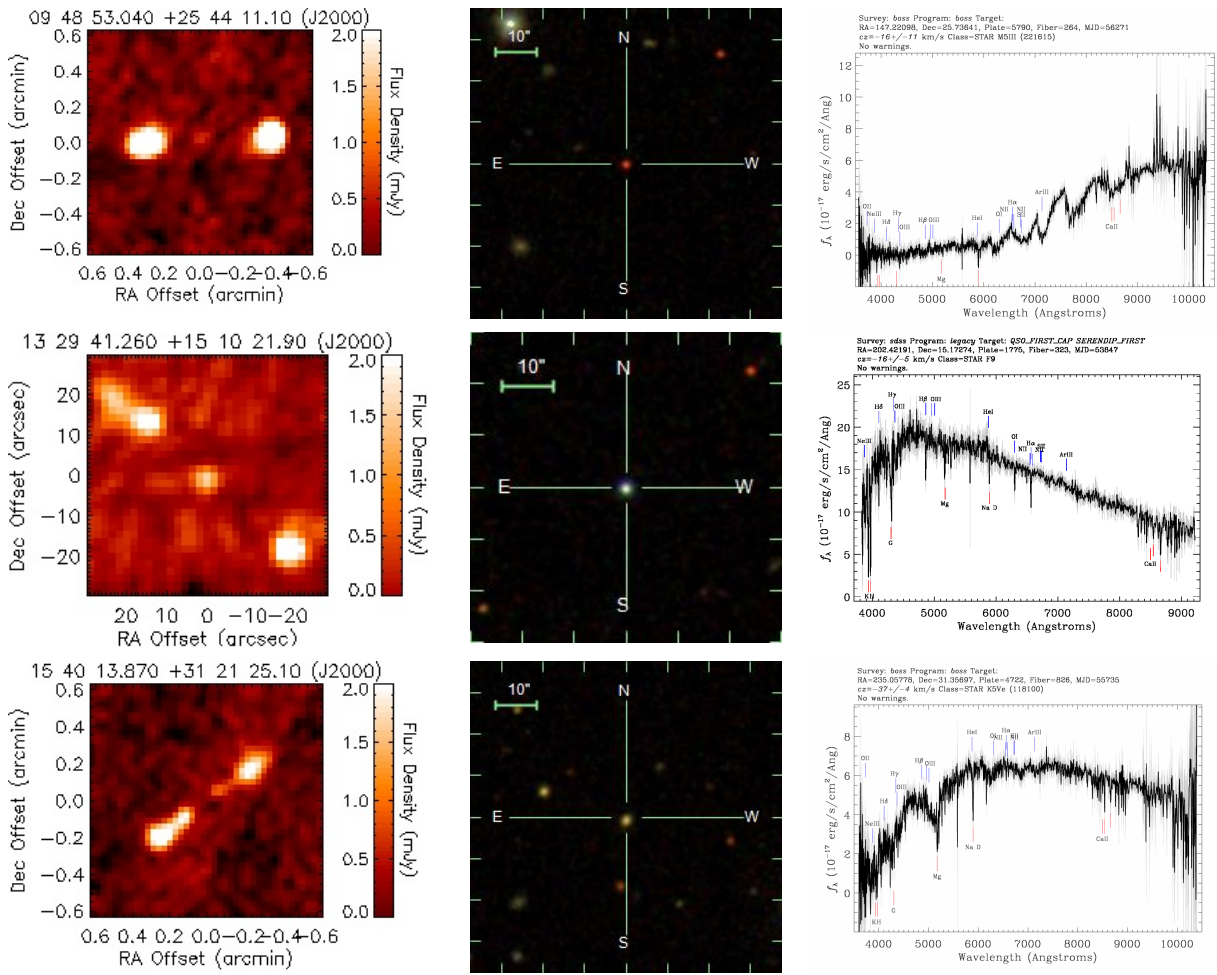

**Fig. 1. Examples of 3 new radio stars with double-lobed radio emission found by us; each row shows the radio image on the left, the optical image at center, and the optical spectrum at right (see text for details)**

We visually inspected the most promising 38 candidates for double-lobed sources, fulfilling the conditions $0.5 \leq$ FLR $\leq 2.0$, $0.5 \leq$ ALR $\leq 2.0$ and DPA $\leq 5°$. We found three convincing examples in which the lobes had no optical or infrared counterpart, and which we present in Fig. 1. The upper row shows SDSS J132941.25+151021.8, an F9 star with a radio core and two symmetrically placed lobes. This star had been reported to have complex radio emission by Kimball et al. (2009) but without mentioning its symmetric triple structure. The middle row shows SDSS J094853.03+254411.0, an M5 star with a faint radio core of ~0.6 mJy and two radio lobes 43" apart, not only symmetrically placed on opposite sides of the star, but being extended along the major source axis. The optical spectrum leaves no doubt about its stellar nature. The bottom row shows the K3 star SDSS J154013.86+312125.0 within ~2´´ of the center of a symmetrical double radio source of size ~45´´. While the optical spectrum leaves no doubt about the stellar nature, the optical image of this object shows some gradient from green in the NW to red color in the SE, and leaves a possibility that this may be a superposition of a brighter star and a fainter galaxy.

B. Spiral Galaxies

We extracted all radio sources from FIRST within 6.0´ around the 675,874 spiral candidates. Running our source code to look for source pairs aligned with the spiral candidates, we reduced the resulting sample to the most promising pairs, with ALR and FLR closest to unity, the smallest DPA, and radio fluxes > 2.5 mJy each. We discarded all cases where one of the radio sources coincided with an optical object from SDSS within 2.4´´. We inspected a total of 115 such source pairs, but did not find any evidence for a relation with the spiral galaxy except for a single case which had been previously reported by Mao et al. (2015). This is the face-on spiral SDSS J164924.01+263502.5 with a classical double radio source known as B2 1647+26A.
Of those spirals with only a single source within 6', about 1290 fell within 1.5" of the SDSS position, and thus are very likely the true radio nuclei of these spirals. We inspected all 70 radio sources with either a deconvolved size of at least 10", or larger than 5" and integrated flux above 10 mJy. All sources were compatible with the radio emission confined within the plane of the galaxies, and thus being due to star formation activity. No single case was found of radio emission indicating a jet base of a classical radio galaxy oriented perpendicular to the galaxy major axis, again confirming the extreme rareness of these objects.

## CONCLUSIONS

We designed an algorithm to search for pairs of radio sources suggesting the emission of radio jets from them, similar to those known in classical radio galaxies, and applied it to unprecedented samples of 878,031 Galactic stars and 675,874 spiral galaxy candidates. We found three new examples of double-lobed radio stars, while we only rediscovered one previously known double source in a spiral galaxy, confirming that the latter are extremely rare.
As an aside, we found that the detection of radio emission matching an SDSS object classified as a spectroscopic star, is an efficient method to find misclassified optical spectra, either being very noisy or having wrongly identified spectral lines.
The period of 5 weeks for this research project did not allow us to inspect more candidates found with our algorithm. We plan to clean these large lists further, e.g. by eliminating all radio lobes having an optical counterpart in SDSS, and/or by including in our analysis the orientation of those suspected radio lobes that are listed as extended nearly along the major axis of the pair of lobes straddling the central object. Moreover, our algorithm may be applied to other existing and future radio surveys or new optical samples like spectroscopic stars or spiral galaxies.

## ACKNOWLEDGEMENTS

We are grateful to M. A. Jiménez Valencia (Univ. de Sonora, Mexico) for sharing his FORTRAN code to search for radio source alignments around arbitrary sky positions.

## BIBLIOGRAPHY

ALAM, S., ALBARETI, F.D., ALLENDE PRIETO, C., ANDERS, F., ANDERSON, S.F., ANDERTON, T., ANDREWS, B.H., ARMENGAUD, E., AUBOURG, É., and 296 coauthors (2015). The Eleventh and Twelfth Data Releases of the Sloan Digital Sky Survey: Final Data from SDSS-III, *Astrophys. J. Suppl. Ser*. 219, Art.12.

BECKER, R.H., WHITE, R.L., HELFAND, D.J., (1995). The FIRST Survey: Faint Images of the Radio Sky at Twenty Centimeters*, Astrophys. J.,* 450, 559-577

BEST, P.N., KAUFFMANN, G., HECKMAN, T.M., IVEZIĆ, Ž., (2005), A sample of radio-loud active galactic nuclei in the Sloan Digital Sky Survey*, Mon. Not. Roy. astron. Soc*. 362, 9-24

HELFAND, D.J., SCHNEE, S., BECKER, R.H., WHITE, R.L., McMAHON, R.G., (1999). The FIRST Unbiased Survey for Radio Stars, *Astron. J.*, 117, 1568-1577

HELFAND, D.J., WHITE, R.L., BECKER, R.H., (2015). The Last of FIRST: The Final Catalog and Source Identifications, *Astrophys. J.,* 801, Art. 26

HUERTAS-COMPANY, M., AGUERRI, J.A.L., BERNARDI, M., MEI, S., SÁNCHEZ ALMEIDA, J., (2001). Revisiting the Hubble sequence in the SDSS DR7 spectroscopic sample: a publicly available Bayesian automated classification, *Astron. Astrophys.* 525, A157

KIMBALL, A.E., KNAPP, G.R., IVEZIĆ, Ž., WEST, A.A., BOCHANSKI, J.J., PLOTKIN, R.M., GORDON, M., (2009), A Sample of Candidate Radio Stars in First and SDSS, *Astrophys. J.,* 701, 535-546

KUMINSKI, E., SHAMIR, L., (2016). Computer-generated visual morphology catalog of ~3,000,000 SDSS galaxies, *Astrophys. J. Suppl. Ser.* 223, Art. 20

LEDLOW, M.J., OWEN, F.N., KEEL, W. C., (1998). An Unusual Radio Galaxy in Abell 428: A Large, Powerful FR I Source in a Disk-dominated Host, *Astrophys. J.,* 495, 227-238

MAO, M.Y., OWEN, F.N., DUFFIN, R., KEEL, W.C., LACY, M., MOMJIAN, E., MORRISON, G., MROCZKOWSKI, T., NEFF, S., NORRIS, R.P., SCHMITT, H., TOY, V., VEILLEUX, S., (2015), J1649+2635: a grand-design spiral with a large double-lobed radio source, *Mon. Not. Roy. astron. Soc*. 446, 4176-4185

MIRABEL, I.F, RODRÍGUEZ, L.F., (1999). Sources of Relativistic Jets in the Galaxy, *Ann. Rev. Astron. & Astroph.* 37, 409-443

SÁNCHEZ ALMEIDA, J., AGUERRI, J.A.L., MUÑOZ, C., DE VICENTE, A., (2010). Automatic Unsupervised Classification of All Sloan Digital Sky Survey Data Release 7 Galaxy Spectra, *Astrophys. J.,* 714, 487-504

SCHMITT, J.H.M.M., SCHRÖDER, K.-P., RAUW, G., HEMPELMANN, A., MITTAG, M., GONZÁLEZ-PÉREZ, J.N., CZESLA, S., WOLTER, U., JACK, D., EENENS, P., TRINIDAD, M.A., (2014). TIGRE: A new robotic spectroscopy telescope at Guanajuato, Mexico, *Astron. Nachr.,* 335, 787-796

SINGH, V., ISHWARA-CHANDRA, C.H., SIEVERS, J., WADADEKAR, Y., HILTON, M., BEELEN, A., (2015), Discovery of rare double-lobe radio galaxies hosted in spiral galaxies, *Mon. Not. Roy. astron. Soc*. 454, 1556-1572

TAYLOR, M., (2013). TOPCAT - Tool for OPerations on Catalogues And Tables, Starlink User Note 253 (see also http://www.star.bris.ac.uk/~mbt/topcat)

WILLETT, K.W., LINTOTT, C.J., BAMFORD, S.P., MASTERS, K.L., SIMMONS, B.D., CASTEELS, K.R.V., EDMONDSON, E.M., FORTSON, L.F., KAVIRAJ, S., KEEL, W.C., MELVIN, T., NICHOL, R.C., RADDICK, M.J., SCHAWINSKI, K., SIMPSON, R.J., SKIBBA, R.A., SMITH, A.M., THOMAS, D., (2013). Galaxy Zoo 2: detailed morphological classifications for 304,122 galaxies from the Sloan Digital Sky Survey, *Mon. Not. Roy. astron. Soc.*, 435, 2835-2860